\pdfoutput=1
\documentclass[aps,prl,twocolumn,superscriptaddress,amsmath,amssymb,floatfix]{revtex4-2}
\usepackage{graphicx}
\usepackage{bm}
\usepackage{amsmath,amssymb}
\usepackage{dcolumn}
\newcommand{\ee}{\bar{\bm e}}
\newcommand{\bxi}{\bm\xi}

\begin{document}

\title{Anomalous Hardening of a Fluctuation-Induced Gap}

\author{Shu He}
\email{heshu1987@foxmail.com}
\affiliation{Department of Physics and Electronic Engineering, Sichuan Normal University, Chengdu 610066, China}
\author{Liwei Duan}
\email{duanlw@gmail.com}
\affiliation{Department of Physics, Zhejiang Normal University, Jinhua 321004, China}

\begin{abstract}
In a mean-field picture, a gap opened by quantum fluctuations grows with them. This can fail if the fluctuating mode is critical and its displacement compensates its coupling to the degenerate states. A single parameter sets the gap response: above a universal threshold, the gap hardens as the critical fluctuations weaken. Whenever the parameter is nonzero, overcompensation beyond the critical point drives a secondary phase transition. In a quantum Rabi ring with synthetic flux, exact diagonalization shows the gap of a photonic Landau level rising by 15\% as the critical variance falls by 25\%.
\end{abstract}

\maketitle

Degenerate classical ground states are often split by quantum fluctuations. In frustrated magnets, the harmonic zero-point energy of spin waves selects an ordered state and gaps the pseudo-Goldstone modes, a mechanism known as order by disorder~\cite{Villain1980}, here in its quantum form~\cite{Shender1982,Henley1989,Rau2018,Khatua2026}. In rapidly rotating gases, trap anharmonicity splits the lowest Landau level~\cite{Fetter2009,Cooper2008}, while interactions can drive crystallization within it~\cite{Mukherjee2022}. Near metallic quantum critical points, low-energy particle--hole excitations can select new orders~\cite{Kruger2012,Green2018}. In these settings the modes that lift the degeneracy are not critical, and in a mean-field (Hartree) picture the gap grows with their fluctuations; at finite temperature competing contributions can reverse this trend~\cite{LinShi2025}. Here the degenerate states couple to a bosonic mode that is itself critical: its fluctuations diverge, and it must be treated as an anharmonic oscillator. What then removes the degeneracy, and how does the gap respond when the critical fluctuations are weakened?

Few-body light--matter systems realize this situation in a controllable setting. In the quantum Rabi model the limit $\eta\to\infty$ of the ratio $\eta$ of atomic to cavity frequency plays the role of the thermodynamic limit~\cite{Hwang2015,Liu2017,Hwang2016}, and the superradiant phase transition, including its multicritical versions, has been simulated experimentally~\cite{Cai2021,Wu2024}. In a ring of three Rabi units coupled by photon hopping with a phase $\theta$ (the quantum Rabi triangle, or ring), the normal phase (NP), the ferromagnetic superradiant phase (FSP), and the chiral superradiant phase (CSP) meet at triple points~\cite{Zhang2021,Padilla2022}, where for $\eta\to\infty$ two excitation energies vanish with different critical exponents~\cite{Padilla2022,Zhao2023}. Critical exponents~\cite{Qin2024} and sensing~\cite{Tang2025} in this system have since been studied, as have multicritical Rabi models~\cite{Ye2025,Lu2024} and triangles with real hopping~\cite{Zhao2022,He2026}. At the triple point, however, the quadratic theory leaves the chiral excitations degenerate. How this degeneracy is lifted at finite $\eta$, and what then sets the gap, has not been addressed.

\begin{figure*}[t]
\includegraphics[width=\textwidth]{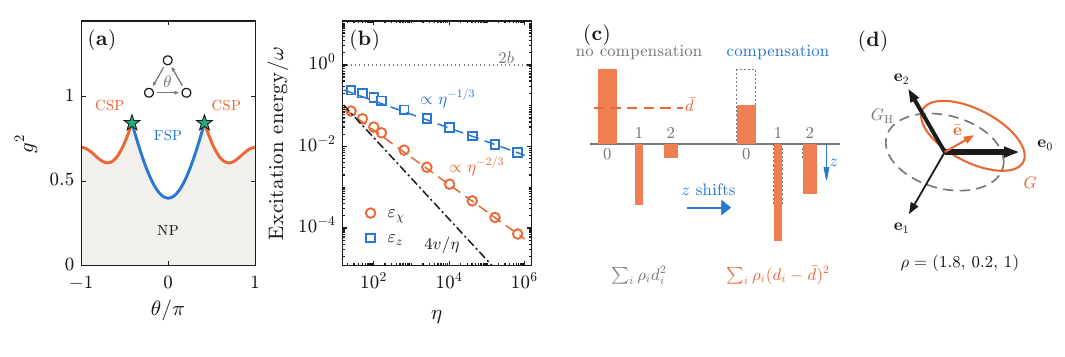}
\caption{Splitting of the photonic Landau level by critical fluctuations, $j=0.6$. (a) Phase boundaries $g^2=A$ (NP--FSP) and $g^2=C-b^2/C$ (NP--CSP); stars, triple points; inset, the ring with hopping phase $\theta$. (b) $\varepsilon_\chi$ and $\varepsilon_z$ for identical atoms from exact diagonalization of Eq.~\eqref{eq:H}; dashed, guides to the eye $\propto\eta^{-2/3}$ and $\eta^{-1/3}$; dash-dotted, the splitting $4v\omega/\eta$ due to the chiral nonlinearity alone; dotted, the cyclotron gap $2\omega b$. (c) Compensation for $\rho=(1.8,0.2,1)$. A guiding-center displacement shifts the site quadratures $x_i$ by $d_i\propto\bm e_i\cdot\bxi$ (bar widths increase with $\rho_i$) at an energy cost $\sum_i\rho_id_i^2$ (left). Shifting $z$ moves every site by $-\bar d$, the weighted mean of the $d_i$ (dashed line), and lowers the cost to $\sum_i\rho_i(d_i-\bar d)^2$ (right; dotted outlines, before the shift). (d) Site vectors $\bm e_i$ (arrow widths increase with $\rho_i$), their weighted mean $\ee$ (orange), and their weighted spread about $\ee$ ($G$, solid) and about the origin ($G_{\rm H}$, dashed).}
\label{fig:1}
\end{figure*}

In this Letter, we show that the degenerate chiral states form a lowest Landau level and that the zero-point fluctuations of the critical uniform mode lift its degeneracy, opening a chiral gap $\propto\eta^{-2/3}$. The uniform mode is displaced together with the guiding center of the Landau level and thereby cancels part of the confinement produced by their coupling [Fig.~\ref{fig:1}(c)]. The gap is therefore set by how the three sites, represented as vectors in the plane of the guiding center and weighted by their atomic nonlinearities, spread about their weighted mean, not about the origin as in a mean-field treatment [Fig.~\ref{fig:1}(d)]. Away from the triple point the critical fluctuations weaken, but so does the compensation, which restores part of the canceled confinement. A single parameter $a$ decides which effect wins: above a universal threshold $a_c\simeq0.83$, the gap hardens while the fluctuations that produce it weaken. On the superradiant side the displacement instead overcompensates, and for any $a>0$ this drives a secondary phase transition. The mechanism requires only a critical mode with quartic anharmonicity coupled to a zero-frequency mode, and exact diagonalization confirms it without adjustable parameters.

\textit{Model and lowest Landau level.}---The ring is described by ($\hbar=1$)
\begin{align}
H=&\sum_{i=0}^{2}\Big[\omega a_i^\dagger a_i+\frac{\Delta_i}{2}\sigma_i^z+\lambda_i(a_i+a_i^\dagger)\sigma_i^x\Big]\nonumber\\
&-J\sum_{i=0}^{2}\big(e^{-\mathrm{i}\theta}a_i^\dagger a_{i+1}+\mathrm{H.c.}\big),
\label{eq:H}
\end{align}
with periodic indices. We write $j=2J/\omega$, $\eta=\Delta/\omega$ for a reference atomic frequency $\Delta$, $\Delta_i=\Delta/\rho_i$ with frequency ratios $\rho_i$, and $g=2\lambda_i/\sqrt{\omega\Delta_i}$, kept equal at all sites by adjusting each $\lambda_i$. For large $\eta$, projecting onto the lower atomic level gives the low-energy effective Hamiltonian of the photons~\cite{Hwang2015}, $H_{\rm eff}=h_2+\sum_ig^4\rho_ix_i^4/(4\eta)$ in units of $\omega$, with $x_i=(a_i+a_i^\dagger)/\sqrt2$ and $h_2$ quadratic. Inhomogeneous atomic frequencies thus change only the quartic coefficients. In real Fourier coordinates,
\begin{align}
x_i=\frac{z}{\sqrt3}+\sqrt{\frac23}\,\big(X\cos\alpha_i+Y\sin\alpha_i\big),\qquad \alpha_i=\frac{2\pi i}{3},
\label{eq:coords}
\end{align}
the quadratic Hamiltonian is
\begin{align}
h_2=&\,\frac{A}{2}p_z^2+\frac{r_0}{2}z^2+\frac{r_\chi}{2}R^2\nonumber\\
&+\frac{C}{2}\Big[\Big(p_X+\frac{b}{C}Y\Big)^2+\Big(p_Y-\frac{b}{C}X\Big)^2\Big],
\label{eq:h2}
\end{align}
with $R^2=X^2+Y^2$, $A=1-j\cos\theta$, $C=1+(j/2)\cos\theta$, $b=(\sqrt3/2)j\sin\theta$, $r_0=A-g^2$, and $r_\chi=C-g^2-b^2/C$. The quadratic uniform and chiral excitation energies vanish at $r_0=0$ (NP--FSP boundary) and $r_\chi=0$ (NP--CSP boundary), respectively; the two boundaries meet at the triple point $\cos\theta_*=(\sqrt{1+2j^2}-1)/(2j)$, $g_*^2=A_*\equiv A(\theta_*)$ [Fig.~\ref{fig:1}(a)].

At $r_\chi=0$ the chiral part of Eq.~\eqref{eq:h2} describes a charged particle in a uniform magnetic field, or an oscillator in a frame rotating at its own frequency, with oscillator length $\ell=(C/b)^{1/2}$. Its two circular modes have frequencies $2\omega b$, the cyclotron gap, and zero. With the cyclotron mode in its ground state, the states of the zero-frequency mode form the degenerate lowest Landau level. The two quadratures of this mode, $\bxi=(Q,P)^T$ with $[Q,P]=\mathrm{i}$, are the coordinates of the guiding center and do not appear in $h_2$~\cite{SM}. At $r_0=0$ the uniform coordinate $z$ has no quadratic restoring force either. The quartic terms couple the two: the zero-point motion of the critical uniform mode, whose amplitude grows as $\eta^{1/6}$, slightly saturates every atom and thereby shifts the energies of the chiral photons. This cross-Kerr shift, akin to vacuum-induced dispersive shifts~\cite{Lolli2015,Lu2018}, exceeds the effect of the chiral nonlinearity because the fluctuations of $z$ diverge. It opens the chiral gap $\varepsilon_\chi$, far below the lowest uniform excitation energy $\varepsilon_z$ [Fig.~\ref{fig:1}(b)].

\begin{figure*}[t]
\includegraphics[width=\textwidth]{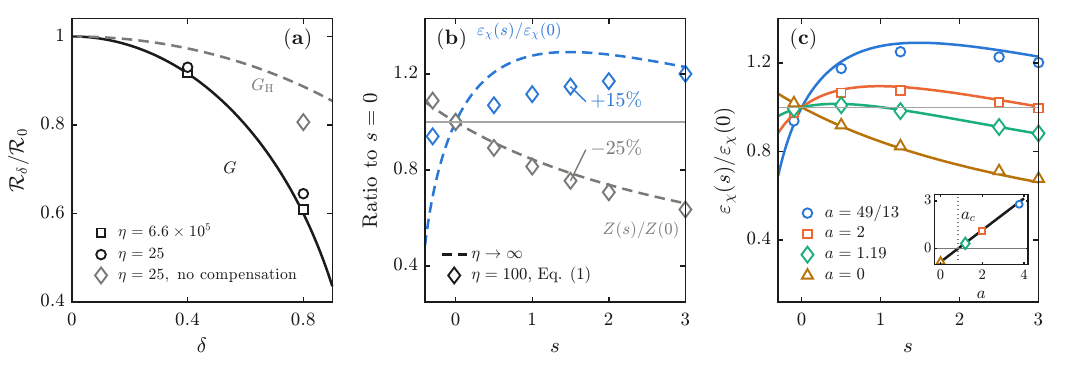}
\caption{Covariance and gap response, $\rho=(1+\delta,1-\delta,1)$, $j=0.6$. (a) $\mathcal R_\delta/\mathcal R_0$ at the triple point from exact diagonalization of Eq.~\eqref{eq:H} for $\eta=25$ and $6.6\times10^5$; solid, from the covariance $G$ [Eq.~\eqref{eq:gap}]; dashed, from the mean-field second moment $G_{\rm H}$; diamond, $H_{\rm eff}$ at $\eta=25$ without the coupling that produces the compensation. (b) $\delta\simeq0.922$ ($a=49/13$): $\varepsilon_\chi(s)/\varepsilon_\chi(0)$ (blue) and $Z(s)/Z(0)$ (gray) at fixed local quartic coefficients; diamonds, exact diagonalization of Eq.~\eqref{eq:H} at $\eta=100$, with the values at $s=1.5$ marked; dashed, $\eta\to\infty$ [Eq.~\eqref{eq:ratio} and $f(s)/f(0)$]; the offset between diamonds and dashed lines is a finite-$\eta$ correction, which for $s<0$ includes the rounding of the secondary transition (Fig.~\ref{fig:3}). (c) Gap response for $\delta=0$, $0.8$, $\sqrt3/2$, and $0.922$ ($a=0$, $1.19$, $2$, and $49/13$), whose leading-order maximum gains for $s\ge0$ are $0$, $1.5\%$, $9.5\%$, and $29\%$: symbols, $H_{\rm eff}$ at $\eta=10^6$; lines, Eq.~\eqref{eq:ratio}. Inset, initial slope $12f(0)\,d\ln\varepsilon_\chi/ds$ against $a$, which changes sign at $a_c$: line, Eq.~\eqref{eq:ac}; symbols, $H_{\rm eff}$ at $\eta=10^8$.}
\label{fig:2}
\end{figure*}

\textit{Hamiltonian of the guiding center.}---The cyclotron motion stays in its ground state when $2b\gg\eta^{-1/3}$, and $x_i\to(z+\ell\,\bm e_i\cdot\bxi)/\sqrt3$, with the site vectors $\bm e_i=(\cos\alpha_i,-\sin\alpha_i)^T$~\cite{SM}. The quartic energy becomes
\begin{align}
V=\frac{g^4}{36\eta}\sum_i\rho_i\big(z+\ell\,\bm e_i\cdot\bxi\big)^4.
\label{eq:V}
\end{align}
Since $\varepsilon_z\propto\eta^{-1/3}$ is much larger than the energy scale of the guiding center, $\bxi$ enters the $z$ problem as a parameter (adiabatic approximation). At the triple point $h_2$ is invariant under shifts of $z$, so the shift $z\to z-\ell\,\ee\cdot\bxi$, with $\ee=\sum_i\rho_i\bm e_i/\sum_i\rho_i$ the weighted mean of the site vectors, removes the term $z^3(\ee\cdot\bxi)$ at no energy cost: the uniform mode is displaced together with the guiding center and compensates this coupling. In terms of the sites, a displacement of the guiding center shifts each $x_i$ by $d_i\propto\bm e_i\cdot\bxi$, the uniform mode moves all sites back by their weighted mean $\bar d$, and only the spread of the $d_i$ about $\bar d$ costs energy [Fig.~\ref{fig:1}(c)]. The remaining term quadratic in $z$ gives, with $Z=\langle z^2\rangle$, the energy of the guiding center
\begin{align}
E(\bxi)=\frac{Z}{2\eta}\,\bxi^TG\,\bxi,\qquad G=\frac{g^4\ell^2}{3}\sum_i\rho_i(\bm e_i-\ee)(\bm e_i-\ee)^T.
\label{eq:E}
\end{align}
The matrix $G$ is the weighted covariance of the site vectors about their mean. A mean-field (Hartree) treatment, which replaces $z^2$ by its average without this shift, gives instead their second moment about the origin, $G_{\rm H}=(g^4\ell^2/3)\sum_i\rho_i\bm e_i\bm e_i^T=G+g^4\ell^2\bar\rho\,\ee\ee^T$ with $\bar\rho=\sum_i\rho_i/3$, as in the parallel-axis theorem [Fig.~\ref{fig:1}(d)]. Equation~\eqref{eq:E} also holds with $\bxi$ an operator~\cite{SM}. Because $Q$ and $P$ are conjugate, $E(\bxi)$ is the Hamiltonian of the guiding center, and its level spacing gives the gap, consistent with the curvature formula of Ref.~\cite{Rau2018}:
\begin{align}
\frac{\varepsilon_\chi}{\omega}=\gamma\,\frac{Z}{\eta},\qquad \gamma=\sqrt{\det G}=\gamma_0\sqrt{\frac{\rho_0\rho_1\rho_2}{\bar\rho}},
\label{eq:gap}
\end{align}
where $\gamma_0=g^4\ell^2/2$ is the value for identical atoms. The quartic oscillator for $z$ gives $Z\propto\eta^{1/3}$, so $\varepsilon_\chi\propto\eta^{-2/3}$, between $\varepsilon_z\propto\eta^{-1/3}$~\cite{Hwang2015,Dusuel2004,Vidal2006} and the splitting $4v\omega/\eta$, $v=g^4\ell^4/8$, due to the chiral nonlinearity alone [Fig.~\ref{fig:1}(b)]~\cite{SM}.

Inhomogeneous atomic frequencies distinguish $G$ from $G_{\rm H}$. For $\rho=(1+\delta,1-\delta,1)$, Eq.~\eqref{eq:gap} gives $\gamma/\gamma_0=\sqrt{1-\delta^2}$, whereas $G_{\rm H}$ would give $\sqrt{1-\delta^2/3}$, larger by $48\%$ at $\delta=0.8$, where $\Delta_1/\Delta_0=9$. In exact diagonalization of Eq.~\eqref{eq:H} with all eight atomic states for $25\le\eta\le6.6\times10^5$, the ratio $\mathcal R_\delta=\eta\varepsilon_\chi/(\omega Z)$ extrapolates to within $0.3\%$ of the value from $G$ [Fig.~\ref{fig:2}(a)]~\cite{SM}. Removing from $H_{\rm eff}$ the coupling that produces the compensation moves $\mathcal R_\delta$ toward the value from $G_{\rm H}$ [diamond in Fig.~\ref{fig:2}(a)].

The compensation removes a rank-one part of the confinement, $G_{\rm H}-G\propto\ee\ee^T$, so its strength is measured by a single parameter,
\begin{align}
1+a=\frac{\det G_{\rm H}}{\det G},\qquad a=g^4\ell^2\bar\rho\;\ee^{\,T}G^{-1}\ee,
\label{eq:a}
\end{align}
which quantifies how much of the mean-field confinement is canceled by the displacement of the uniform mode. It vanishes only for identical atoms, diverges when one atom loses its nonlinearity, and equals $2\delta^2/[3(1-\delta^2)]$ for $\rho=(1+\delta,1-\delta,1)$~\cite{SM}.

\begin{figure}[t]
\includegraphics[width=\columnwidth]{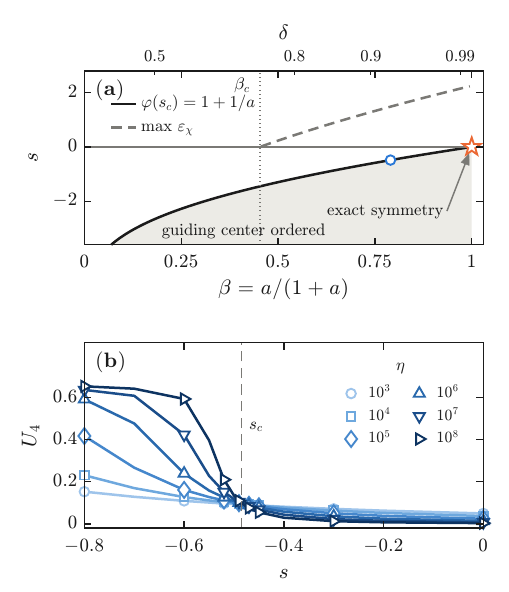}
\caption{Overcompensation and the secondary phase transition. (a) Phase diagram in $s$ and $\beta=a/(1+a)$, which maps $0\le a<\infty$ onto $0\le\beta<1$: solid, critical line $s_c$ of the secondary transition, Eq.~\eqref{eq:sc}, which for all cases of the Rabi ring examined is of second order and bounds the ordered (shaded) phase; dashed, maximum of the gap for $\beta>\beta_c=a_c/(1+a_c)$ (dotted line); circle, Binder-cumulant crossing of (b) for $\eta=10^7$ and $10^8$; star ($\beta=1$), one atom without nonlinearity, where the line ends at the triple point in an exact symmetry. Top axis, $\delta$ for $\rho=(1+\delta,1-\delta,1)$. (b) Binder cumulant $U_4$ of the guiding-center coordinate along the unstable direction for $\delta\simeq0.922$, from $H_{\rm eff}$ at $10^3\le\eta\le10^8$; dashed, $s_c$.}
\label{fig:3}
\end{figure}

\textit{Weaker fluctuations, larger gap.}---Moving away from the triple point along the NP--CSP boundary ($r_\chi=0$) by $r_0>0$, at fixed local quartic coefficients, gives $z$ a quadratic restoring force. In terms of the scaled distance $s\propto r_0\eta^{2/3}$ and coordinate $y\propto\eta^{-1/6}z$~\cite{SM}, the $z$ problem becomes $p_y^2+y^4+sy^2$, whose ground-state variance $f(s)=\langle y^2\rangle$ decreases monotonically with $s$; $Z\propto f(s)$ measures the critical fluctuations. The second-order energy of the cubic term, which produces the compensation, becomes $\varphi(s)$ times the energy of the shift, while the term quadratic in both $z$ and $\bxi$, which gives $G_{\rm H}$, is unchanged. An energy-weighted sum rule makes this compensation ratio exactly $\varphi(0)=1$ at the triple point, where the shift cancels the whole difference $G_{\rm H}-G$, and it decays as $(11/3)s^{-3/2}$ for $s\gg1$~\cite{SM}. Hence, with $Z$ evaluated at $s$, the confinement is the mean-field one minus $\varphi(s)$ times this difference,
\begin{align}
G(s)=G_{\rm H}-\varphi(s)\big(G_{\rm H}-G\big).
\label{eq:Gs}
\end{align}
Because $G_{\rm H}-G$ has rank one, Eqs.~\eqref{eq:gap}--\eqref{eq:Gs} give the gap ratio,
\begin{align}
\frac{\varepsilon_\chi(s)}{\varepsilon_\chi(0)}=\frac{f(s)}{f(0)}\sqrt{1+a\big[1-\varphi(s)\big]}.
\label{eq:ratio}
\end{align}
The two factors compete. The first is the weakening of the critical fluctuations; alone, it would make the gap proportional to $Z$ at leading order, as in a mean-field picture and as for identical atoms. The second is the confinement restored as the compensation weakens. Expanding $\varphi$ and $f$ to first order in $s$~\cite{SM},
\begin{align}
\left.\frac{d\ln\varepsilon_\chi}{ds}\right|_{s=0}=\frac{a-a_c}{12f(0)},\qquad a_c=12\,|f'(0)|=0.8282,
\label{eq:ac}
\end{align}
where $f(0)=0.3620$ and $f'(0)=-0.0690$ are properties of the quartic oscillator alone. For $a>a_c$ the gap therefore hardens as the critical fluctuations weaken [inset of Fig.~\ref{fig:2}(c)]; since $\varphi\to0$ and $f\propto s^{-1/2}$ at large $s$, it then passes through a maximum [dashed line in Fig.~\ref{fig:3}(a)]. For $\rho=(1+\delta,1-\delta,1)$ with $\delta\simeq0.922$, where $a=49/13\simeq3.77$ and $\Delta_1/\Delta_0\simeq25$, the gap rises by up to $29\%$, at $s=1.48$, where $Z$ has fallen by $21\%$. With two identical atoms, $a>a_c$ requires the frequency of the third atom to differ from theirs by a factor of more than 5.5~\cite{SM}.

Exact diagonalization of Eq.~\eqref{eq:H} shows the effect already at $\eta=100$ [Fig.~\ref{fig:2}(b)]: for $\delta\simeq0.922$ and $s=1.5$ the gap rises by $15\%$ while $Z$ falls by $25\%$, whereas for identical atoms the gap falls by $6\%$. The remaining difference from Eq.~\eqref{eq:ratio} is a finite-$\eta$ correction [Fig.~\ref{fig:2}(c)]: for $s\ge0$, $H_{\rm eff}$ approaches Eq.~\eqref{eq:ratio} for all four values of $\delta$ tested, agreeing to within $1\%$ at $\eta=10^8$, and the gap--variance ratio of the Hamiltonian projected onto the lowest Landau level agrees with $\sqrt{\det G(s)}$ to the same accuracy~\cite{SM}. Equation~\eqref{eq:ratio} describes the response to $r_0$ alone, at fixed local quartic coefficients, so these calculations change $j$ and $\theta$ at fixed $g$ along $r_\chi=0$; $A$ and $\ell^2$ then change only slightly. Tuning $g$ at fixed $\theta$ instead also changes $g^4$ and $\ell^2$ and thus the nonuniversal prefactor of the gap, at relative order $s\eta^{-2/3}$. This change vanishes as $\eta\to\infty$, but its large coefficients mask the effect at $\eta=100$~\cite{SM}.

\textit{Overcompensation and a secondary phase transition.}---For $s<0$, on the superradiant side, the uniform potential develops two shallow wells and $\varphi(s)>1$: the displacement of $z$ overcompensates the coupling, and Eq.~\eqref{eq:Gs} gives a confinement below $G$. Since $\det G(s)\propto1+a[1-\varphi(s)]$ and $\varphi$ grows without bound as $s\to-\infty$~\cite{SM}, the confinement of the guiding center vanishes at
\begin{align}
\varphi(s_c)=1+\frac1a
\label{eq:sc}
\end{align}
for every $a>0$, unlike the hardening, which at leading order requires $a>a_c$; for $a=49/13$, $s_c=-0.485$ [Fig.~\ref{fig:3}(a)]. For $s<s_c$ the state with $\langle\bxi\rangle=0$ is unstable.

A sharp transition follows because the guiding center becomes classical. Measured in units of the quartic length $\propto\eta^{1/6}$ of $z$, its two quadratures have the commutator $\mathrm{i}\hbar_{\rm eff}$, with an effective Planck constant $\hbar_{\rm eff}\propto\eta^{-1/3}$. As $\eta\to\infty$ the guiding center therefore reaches its classical limit, as the cavity mode does in the Rabi model~\cite{Hwang2015}, while the uniform mode remains quantum, and it sits at the minimum of the adiabatic potential $V_{\rm ad}$, the ground-state energy of the $z$ problem at fixed guiding center. Because $V_{\rm ad}$ is even, its minimum must leave the origin at some $s\ge s_c$: a secondary phase transition that breaks the $\mathbb Z_2$ parity and is defined for $\eta\to\infty$ at fixed $s$, that is, within the critical region $r_0\propto\eta^{-2/3}$. For all $\rho$ of the Rabi ring examined, it is of second order, occurs at $s_c$, and lies in the universality class of the Rabi model, so that the gap at $s_c$ scales as $\eta^{-7/9}$~\cite{SM}. The Binder cumulant of the guiding-center coordinate along the unstable direction sharpens accordingly, with crossings that approach $s_c$ [Fig.~\ref{fig:3}(b)].

As $a\to\infty$, Eq.~\eqref{eq:sc} gives $s_c\to0^-$, and the critical line reaches the triple point [Fig.~\ref{fig:3}(a)]. This limit is realized when one atom has no nonlinearity. In this limit, a magnetic translation of $(X,Y)$ combined with a shift of $z$ commutes with Eq.~\eqref{eq:H} at the triple point, so the energy of the guiding center is exactly flat along one direction~\cite{SM}.

\textit{Generality and outlook.}---The response of the gap and the instability depend on the ring only through a general effective Hamiltonian, in which a critical mode with quartic anharmonicity couples through quartic terms to a zero-frequency mode~\cite{SM}. Whenever the state $\langle\bxi\rangle=0$ is stable at the critical point, Eqs.~\eqref{eq:ratio}--\eqref{eq:sc} hold with $a$ fixed by the couplings; if the secondary transition is of first order, Eq.~\eqref{eq:sc} gives its spinodal. We have verified this for random couplings without any ring structure~\cite{SM}. The zero-frequency mode can also arise from resonance rather than flux: for a cavity mode at zero detuning in the frame of a parametrically driven Kerr oscillator at threshold, $a/(1+a)$ is the share of the critical Kerr nonlinearity carried by a Kerr element shared by the two modes, and $a=49/13$ again gives a $29\%$ rise of the gap~\cite{SM}. The coupled-top triangle of Ref.~\cite{Duan2023} with an added Dzyaloshinskii--Moriya interaction also realizes this Hamiltonian, with $a=0$ for identical sites~\cite{SM}. So does a particle in a trap rotating at its radial frequency, with quartic anharmonicity and no axial confinement~\cite{Fetter2009}, where $a=0$ if the anharmonicity is axially symmetric~\cite{SM}; rotating gases have reached the lowest Landau level and squeezed its guiding center~\cite{Fletcher2021}. Unlike photonic Landau levels engineered in real space~\cite{Schine2016,Ozawa2019} or coupled to a cavity~\cite{Mivehvar2026}, the Landau level here lives in the quadrature plane of two ring modes.

Trapped ions have reached $\eta=25$ for a single Rabi model~\cite{Cai2021} and engineered Peierls phases on motional couplings~\cite{Kiefer2019}, and superconducting circuits have realized a three-site ring with synthetic flux~\cite{Roushan2017}. The gap and the critical variance are both measurable, $Z$ as the variance of the uniform quadrature $(x_0+x_1+x_2)/\sqrt3$ and $\varepsilon_\chi$ as the response frequency of $X+\mathrm{i}Y$, and their ratios are free of $\eta$ and of fit parameters at leading order. Already at $\eta=25$ the ratio $\mathcal R_\delta/\mathcal R_0$ distinguishes $G$ from $G_{\rm H}$, and the $15\%$ rise at $\eta=100$ occurs at $r_0\simeq0.016$; both require $r_0$ and $r_\chi$ to be set to within about $10^{-3}$. Here $\eta$ refers to the reference atom; the most weakly nonlinear atom has the larger ratio $\Delta_1/\omega=\eta/\rho_1$, about $125$ for $\delta=0.8$ at $\eta=25$ and $1.3\times10^3$ for $\delta\simeq0.922$ at $\eta=100$.

In summary, when a critical mode with quartic anharmonicity is displaced together with a zero-frequency mode, the gap it opens responds to the critical fluctuations through a single parameter $a$. Above $a_c\simeq0.83$ the gap hardens as the fluctuations weaken, and for any $a>0$ overcompensation on the superradiant side drives a secondary phase transition in which the zero-frequency mode reaches its classical limit. In the quantum Rabi ring this mechanism splits a photonic Landau level, and it carries over to every system described by the same general effective Hamiltonian.

\textit{Data availability.}---Code and data are available at Zenodo, doi:10.5281/zenodo.24488626.

This work was supported by the National Natural Science Foundation of China (Grants No.~11804240 and No.~12305032). OpenAI ChatGPT/Codex assisted with language editing, code packaging, and cross-checks of derivations and numerics; the authors verified all results and are responsible for the content.

\clearpage
\onecolumngrid
\setcounter{equation}{0}\setcounter{figure}{0}\setcounter{table}{0}
\renewcommand{\theequation}{S\arabic{equation}}
\renewcommand{\thefigure}{S\arabic{figure}}
\renewcommand{\thetable}{S\arabic{table}}

\begin{center}
{\large\bfseries Supplemental Material for ``Anomalous Hardening of a Fluctuation-Induced Gap''}\\[4pt]
Shu He and Liwei Duan
\end{center}

\noindent Energies are in units of $\omega$ ($\Omega$ for the spin triangle of Sec.~S6), and equation numbers without the prefix S refer to the main text. Section~S1 derives the effective Hamiltonian and the lowest Landau level [Eqs.~\eqref{eq:H}--\eqref{eq:V}]; S2 treats identical atoms, the scaled uniform oscillator, and the first correction in $\eta^{-1/3}$; S3 the shift of the uniform mode, the sum rule, the compensation ratio, and the parameter $a$ [Eqs.~\eqref{eq:E}--\eqref{eq:Gs}]; S4 the response of the gap [Eqs.~\eqref{eq:ratio} and \eqref{eq:ac}]; S5 the secondary phase transition [Eq.~\eqref{eq:sc}]; S6 the general effective Hamiltonian, its test on random couplings, and other realizations; S7 one atom without nonlinearity; and S8 the numerical methods. Complete numerical tables, including those summarized in the text below, are in the Zenodo archive given in the main text.

\section*{S1. EFFECTIVE HAMILTONIAN AND LOWEST LANDAU LEVEL}

With $\eta_i=\Delta_i/\omega=\eta/\rho_i$ and $g=2\lambda_i/\sqrt{\omega\Delta_i}$, the local rotation $U_i=\exp[-\mathrm{i}\vartheta_i(x_i)\sigma_i^y/2]$, $\tan\vartheta_i=\sqrt2\,g\,x_i/\sqrt{\eta_i}$, diagonalizes the atomic part of site $i$ at fixed $x_i=(a_i+a_i^\dagger)/\sqrt2$. Its lower eigenvalue is
\begin{align}
-\frac{\eta_i}{2}\sqrt{1+\frac{2g^2x_i^2}{\eta_i}}=-\frac{\eta_i}{2}-\frac{g^2x_i^2}{2}+\frac{g^4\rho_ix_i^4}{4\eta}-\frac{g^6\rho_i^2x_i^6}{4\eta^2}+\cdots.
\label{eq:S1}
\end{align}
The quadratic term is the same at every site; the frequency ratios $\rho_i$ enter first at quartic order. The momentum terms generated by $U_i$ give a state-independent $O(\eta^{-1})$ constant, as for a single Rabi model~\cite{Hwang2015}. At the orders considered here, they do not couple Landau-level states.

The real Fourier coordinates of Eq.~\eqref{eq:coords}, with the same transformation for the momenta, diagonalize the hopping. The modes at $k=\pm2\pi/3$ have frequencies $1-j\cos(k-\theta)=C\mp b$, and the atomic term $-g^2\sum_ix_i^2/2=-g^2(z^2+R^2)/2$ completes Eq.~\eqref{eq:h2}. The NP--FSP and NP--CSP boundaries meet where $A=C-b^2/C$, which gives the triple point $\cos\theta_*$ of the main text. For $j=0.6$, $A_*=0.8442561$, $C_*=1.0778719$, $b_*=0.5018046$, and $\ell^2=C_*/b_*=2.1479912$.

\textit{Lowest Landau level.}---At $r_\chi=0$, the circular operators $c_\pm=(a_X\mp\mathrm{i}a_Y)/\sqrt2$, built from the annihilation operators $a_X$, $a_Y$ of oscillators of length $\ell$ in $X$ and $Y$, give $L_z=n_+-n_-$ and the chiral Hamiltonian $b(1+2n_-)$. The states with $n_-=0$ are degenerate in $n_+=n$,
\begin{align}
\psi_n(X,Y)=\frac{1}{\sqrt{\pi n!}\,\ell}\Big(\frac{X+\mathrm{i}Y}{\ell}\Big)^ne^{-R^2/2\ell^2},
\end{align}
and $W=X+\mathrm{i}Y=\ell(c_-+c_+^\dagger)$. Projecting onto this level with $\mathcal P$,
\begin{align}
\mathcal PX\mathcal P=X_c=\frac{\ell}{\sqrt2}Q,\qquad \mathcal PY\mathcal P=Y_c=-\frac{\ell}{\sqrt2}P,\qquad Q=\frac{c_++c_+^\dagger}{\sqrt2},\qquad P=\frac{c_+-c_+^\dagger}{\mathrm{i}\sqrt2},
\end{align}
so that $[Q,P]=\mathrm{i}$ and $[X_c,Y_c]=-\mathrm{i}\ell^2/2$; the conventional magnetic length squared is $\ell^2/2$. Products are projected after multiplication, and the cyclotron vacuum adds constants, for example
\begin{align}
\mathcal PR^2\mathcal P=X_c^2+Y_c^2+\frac{\ell^2}{2}=\ell^2(n+1),\qquad \mathcal PR^4\mathcal P=\ell^4(n+1)(n+2).
\label{eq:S4}
\end{align}
These constants add an $n$-independent term to the coefficient of $z^2$, which shifts $r_0$ by $O(\eta^{-1})$; they enter the first correction but not the leading gap. The cyclotron gap $2b$ exceeds the uniform quartic scale $\eta^{-1/3}$ when $b\eta^{1/3}\gg1$, that is, $\eta\gg8$ at $j=0.6$. The projected local coordinate is $x_i\to(z+\ell\,\bm e_i\cdot\bxi)/\sqrt3$ with $\bxi=(Q,P)^T$ and $\bm e_i=(\cos\alpha_i,-\sin\alpha_i)^T$.

\section*{S2. IDENTICAL ATOMS: ENERGY SCALES AND FIRST CORRECTION}

For $\rho_i=1$, the diagonal part of the projected Hamiltonian in Landau-level state $n$ is
\begin{align}
h_n=\frac{A_*}{2}p_z^2+\Big[\frac{r_0}{2}+\frac{\gamma_0(n+1)}{\eta}\Big]z^2+\frac{u}{\eta}z^4+\mu n+\frac{v}{\eta}(n+1)(n+2),
\label{eq:S5}
\end{align}
with $u=A_*^2/12$, $\gamma_0=A_*^2\ell^2/2$, $v=A_*^2\ell^4/8$, and $\mu=\ell^2r_\chi/2$ (at the triple point $g^4=A_*^2$). The last term is the chiral nonlinearity alone; it splits $n=0$ and $n=1$ by $4v/\eta$. The angular term $z\,\mathrm{Re}\,W^3$ changes $n$ by three and reverses the parity of $z$; it first affects the gap at $O(\eta^{-4/3})$. With $z=\ell_z\eta^{1/6}y$, $\ell_z=(A_*/2u)^{1/6}$, and $M=\ell_z^4/A_*$, the uniform problem becomes
\begin{align}
\Big(-\frac{d^2}{dy^2}+y^4+sy^2\Big)|m\rangle=\mathcal E_m(s)|m\rangle,\qquad s=Mr_0\eta^{2/3},\qquad f(s)=\langle0|y^2|0\rangle=\mathcal E_0'(s),
\label{eq:S6}
\end{align}
with $f(0)=0.3620226$ and $f'(0)=-0.0690205$. Each Landau-level quantum shifts $s$ by $2\gamma_0M\eta^{-1/3}$, and expanding $\mathcal E_0$ gives
\begin{align}
\frac{\varepsilon_\chi}{\omega}&=\mu+\gamma_0\ell_z^2\eta^{-2/3}f(s)+v\eta^{-1}[4+36f'(s)]+O(\eta^{-4/3}),\\
Z&=\ell_z^2\eta^{1/3}f(s)+2\gamma_0M\ell_z^2f'(s)+O(\eta^{-1/3}),\\
\mathcal R-\frac{\eta\mu}{Z}&=\frac{\eta(\varepsilon_\chi/\omega-\mu)}{Z}=\gamma_0+\zeta_1(s)\,\eta^{-1/3}+O(\eta^{-2/3}),\qquad \zeta_1(s)=\frac{v[4+12f'(s)]}{\ell_z^2f(s)}.
\end{align}
The criticality of $z$ sets $Z\propto\eta^{1/3}$ and hence the hierarchy $\eta^{-1/3}\gg\eta^{-2/3}\gg\eta^{-1}$ of $\varepsilon_z$, the chiral gap $\varepsilon_\chi$, and the splitting $4v/\eta$ due to the chiral nonlinearity alone. For identical atoms the ratio of the chiral gap to this splitting is $Z/\ell^2$; at $j=0.6$ it is $0.78$ at $\eta=25$ and grows as $\eta^{1/3}$, reaching about 3 at $\eta=10^3$ and 10 near $\eta=3\times10^4$. In this case the common factor $f(s)$ cancels in $\mathcal R$ at leading order, so Eq.~\eqref{eq:gap} holds at every fixed $s$, and its leading-order form contains only measurable quantities, $\varepsilon_\chi=2(4\lambda^4/\Delta^3)\langle z^2\rangle\langle R^2\rangle$, where $4\lambda^4/\Delta^3$ is the coefficient of $x_i^4$ in the atomic energy. For $j=0.6$ and $s=\mu=0$,
\begin{align}
\frac{\varepsilon_\chi}{\omega}=0.5328\,\eta^{-2/3}+0.6229\,\eta^{-1},\qquad Z=0.6960\,\eta^{1/3}+\cdots,\qquad \mathcal R=0.7655+1.873\,\eta^{-1/3},
\end{align}
so that $\zeta_1(0)=1.873$ and $\zeta_1(0)/\gamma_0=2.447$. Table~\ref{tab:S1} compares $\mathcal R$ with exact diagonalization of Eq.~\eqref{eq:H} up to $\eta=10^{10}$; Fig.~\ref{fig:Scov}(a) shows the approach for $25\le\eta\le655360$. For $s=1,3,10$, $\zeta_1(s)=2.413$, $3.314$, and $5.532$, and $(\mathcal R-\gamma_0)\eta^{1/3}$ at $\eta=10^{10}$ agrees with these values to better than $0.03\%$.

\begin{table}[h]
\caption{Identical atoms, $j=0.6$, $s=\mu=0$: exact diagonalization of Eq.~\eqref{eq:H} at large $\eta$ and the first correction, to be compared with $\zeta_1(0)=1.873$.}
\label{tab:S1}
\begin{ruledtabular}
\begin{tabular}{ccccc}
$\eta$ & $\varepsilon_\chi/\omega$ & $Z$ & $\mathcal R$ & $(\mathcal R-\gamma_0)\eta^{1/3}$\\ \hline
$10^7$ & $1.1541\times10^{-5}$ & 149.07 & 0.77421 & 1.8743\\
$10^8$ & $2.4793\times10^{-6}$ & 322.18 & 0.76955 & 1.8737\\
$10^9$ & $5.3344\times10^{-7}$ & 695.14 & 0.76738 & 1.8734\\
$10^{10}$ & $1.1485\times10^{-7}$ & 1498.7 & 0.76638 & 1.8733
\end{tabular}
\end{ruledtabular}
\end{table}

\section*{S3. DISPLACEMENT OF THE UNIFORM MODE, PARALLEL-AXIS THEOREM, AND COMPENSATION RATIO}

\textit{Shift of $z$.}---Expanding Eq.~\eqref{eq:V} gives
\begin{align}
V=\frac1\eta\Big[uz^4+\frac{g^4\ell\bar\rho}{3}z^3\,\ee\cdot\bxi+\frac{g^4\ell^2}{6}z^2\sum_i\rho_i(\bm e_i\cdot\bxi)^2+O(z\xi^3)\Big],\qquad u=\frac{g^4\bar\rho}{12},
\label{eq:S11}
\end{align}
with $\ee=\sum_i\rho_i\bm e_i/\sum_i\rho_i$. With $\bxi$ treated as a parameter and $r_0=0$, the shift $z\to z-\ell\,\ee\cdot\bxi$ is unitary for the $z$ problem and turns the bracket into $uz^4+(g^4\ell^2/6)z^2\sum_i\rho_i[(\bm e_i-\ee)\cdot\bxi]^2+O(z\xi^3)$. Averaging $z^2$ in the ground state gives Eq.~\eqref{eq:E}. As an operator statement, the canonical change to $z'=z+\ell\,\ee\cdot\bxi$, with the conjugate shift of $\bxi$ by $p_z$, removes the $z^3(\ee\cdot\bxi)$ term and generates momentum terms of relative order $\eta^{-1/3}$.

\textit{General local nonlinearity.}---Let $V=\sum_i\rho_iF(x_i)$ with $x_i=(z+\ell\,\bm e_i\cdot\bxi)/\sqrt3$ and any smooth $F$. After the same shift, $V=\sum_i\rho_iF\big((z+\ell(\bm e_i-\ee)\cdot\bxi)/\sqrt3\big)$. The term linear in $\bxi$ is $(\ell/\sqrt3)F'(z/\sqrt3)\sum_i\rho_i(\bm e_i-\ee)\cdot\bxi$, which vanishes identically, and the quadratic term is $(\ell^2/6)F''(z/\sqrt3)\sum_i\rho_i[(\bm e_i-\ee)\cdot\bxi]^2$. The guiding-center energy is therefore
\begin{align}
E(\bxi)=\frac{\ell^2}{6}\big\langle F''(z/\sqrt3)\big\rangle\sum_i\rho_i\big[(\bm e_i-\ee)\cdot\bxi\big]^2,
\end{align}
which for $F=g^4x^4/(4\eta)$, $\langle F''\rangle=g^4Z/\eta$, is Eq.~\eqref{eq:E}. The derivation requires only that the $z$ Hamiltonian without $V$ be invariant under shifts of $z$, which holds at $r_0=0$, and that the sites differ only by the factors $\rho_i$. In the Rabi ring the sextic terms of Eq.~\eqref{eq:S1} carry factors $\rho_i^2$; they change the covariance at relative order $\eta^{-2/3}$.

\textit{Variance at fixed guiding center and total variance.}---In the derivation of Eq.~\eqref{eq:E}, the variance that enters is that of $z$ at fixed $\bxi$, that is, of the shifted coordinate $z+\ell\,\ee\cdot\bxi$. The total variance $\langle z^2\rangle$, used as $Z$ in the main text, exceeds it by $\ell^2\ee^{\,T}\langle\bxi\bxi^T\rangle\ee=O(1)$, a relative correction of order $\eta^{-1/3}$ that is part of the first correction; for identical atoms $\ee=0$ and the two coincide. When one atom has almost no nonlinearity, the difference is not small (Sec.~S7).

\textit{Sum rule.}---The same result follows from second-order perturbation theory in the cubic term, with $\bxi$ an operator. For $h_0=p_y^2+y^4$, the commutators $[h_0,p_y]=4\mathrm{i}y^3$ and $[p_y,[h_0,p_y]]=12y^2$ give $\langle m|y^3|0\rangle=(\mathcal E_m-\mathcal E_0)\langle m|p_y|0\rangle/(4\mathrm{i})$ and $\sum_m(\mathcal E_m-\mathcal E_0)|\langle m|p_y|0\rangle|^2=6\langle y^2\rangle$, hence
\begin{align}
\mathcal X_3(0)\equiv2\sum_{m>0}\frac{|\langle m|y^3|0\rangle|^2}{\mathcal E_m-\mathcal E_0}=\frac34\langle y^2\rangle.
\label{eq:S13}
\end{align}
For $h_z=(A/2)p_z^2+uz^4/\eta$, with eigenvalues $E_m$, this reads $\sum_{m>0}|\langle m|z^3|0\rangle|^2/(E_m-E_0)=3\eta Z/(8u)$. The second-order energy of $[g^4\ell\bar\rho/(3\eta)]z^3\,\ee\cdot\bxi$ is then $-[g^4\ell^2\bar\rho Z/(2\eta)](\ee\cdot\bxi)^2$, which is the energy of the shift.

\textit{Compensation ratio.}---Away from the triple point ($s\neq0$ in Eq.~\eqref{eq:S6}), we define $\mathcal X_3(s)=2\sum_{m>0}|\langle m|y^3|0\rangle|^2/(\mathcal E_m-\mathcal E_0)$, with $|m\rangle$ and $\mathcal E_m$ the eigenstates and eigenvalues of Eq.~\eqref{eq:S6}, and the compensation ratio
\begin{align}
\varphi(s)=\frac{\mathcal X_3(s)}{\tfrac34f(s)}.
\label{eq:S14}
\end{align}
In scaled variables the second-order energy of the cubic term is $\varphi(s)$ times the energy of the shift at the same $s$, while the $z^2\xi^2$ term of Eq.~\eqref{eq:S11} is unchanged; with $Z$ evaluated at that $s$, this gives Eq.~\eqref{eq:Gs}. For $h_s=p_y^2+y^4+sy^2$, the commutator $[h_s,p_y]=\mathrm{i}(4y^3+2sy)$ gives
\begin{align}
\mathcal X_3(s)=\tfrac34f(s)-\tfrac{s}{8}+\tfrac{s^2}{4}\mathcal X_1(s),\qquad \mathcal X_1=2\sum_{m>0}\frac{|\langle m|y|0\rangle|^2}{\mathcal E_m-\mathcal E_0},
\label{eq:S14b}
\end{align}
so $\varphi(0)=1$ exactly and $\varphi'(0)=-1/[6f(0)]$. For $s\gg1$ the oscillator is harmonic with frequency $2\sqrt s$; then $\mathcal X_3=11/(8s^2)$, $f=1/(2\sqrt s)$, and $\varphi\simeq(11/3)s^{-3/2}$. Table~\ref{tab:S2} lists $\varphi(s)$, obtained by diagonalizing Eq.~\eqref{eq:S6} on a grid; Eq.~\eqref{eq:S14b} holds to all quoted digits. For identical atoms $\ee=0$ and $G(s)=G$.

\begin{table}[h]
\caption{Variance $f(s)$ and compensation ratio $\varphi(s)$ of the quartic oscillator, Eq.~\eqref{eq:S6}.}
\label{tab:S2}
\begin{ruledtabular}
\begin{tabular}{ccccccccccc}
$s$ & 0 & 0.5 & 1 & 2 & 3 & 5 & 10 & 20 & 30 & 100\\ \hline
$f(s)$ & 0.36202 & 0.33103 & 0.30581 & 0.26735 & 0.23939 & 0.20134 & 0.15148 & 0.11001 & 0.09048 & 0.04993\\
$\varphi(s)$ & 1 & 0.8037 & 0.6595 & 0.4672 & 0.3491 & 0.2180 & 0.0975 & 0.0384 & 0.0215 & 0.00364
\end{tabular}
\end{ruledtabular}
\end{table}

\textit{The parameter $a$.}---Because $G_{\rm H}-G=g^4\ell^2\bar\rho\,\ee\ee^{\,T}$ (the parallel-axis theorem) has rank one, $\det[G+t(G_{\rm H}-G)]=\det G\,(1+ta)$ for all $t$, with $a=g^4\ell^2\bar\rho\,\ee^{\,T}G^{-1}\ee=\det G_{\rm H}/\det G-1$. For $\rho=(1+\delta,1-\delta,1)$, $\det G\propto1-\delta^2$ and $\det G_{\rm H}\propto1-\delta^2/3$, so $a=2\delta^2/[3(1-\delta^2)]$. In general, $\det\sum_i\rho_i\bm e_i\bm e_i^T=\tfrac34(\rho_0\rho_1+\rho_1\rho_2+\rho_2\rho_0)$ and Eq.~\eqref{eq:S16} below give $1+a=(\rho_0+\rho_1+\rho_2)(\rho_0\rho_1+\rho_1\rho_2+\rho_2\rho_0)/(9\rho_0\rho_1\rho_2)$, which exceeds 1 by the inequality of arithmetic and geometric means unless all atoms are identical. For two identical atoms, $\rho=(1,1,\rho_2)$, $a>a_c$ requires $\rho_2<0.180$ or $\rho_2>5.55$, and $a=2$ and $49/13$ are reached at $\rho_2=0.092$ and $0.053$. For the general effective Hamiltonian, $a/(1+a)=[3/(4u)]\,\bm c^TB^{-1}\bm c$ (Sec.~S6).

\begin{figure}[h]
\includegraphics[width=0.693\textwidth]{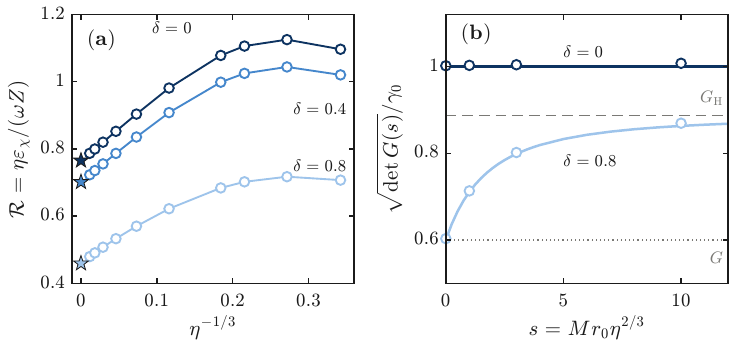}
\caption{Inhomogeneous atomic frequencies at the triple point, $\rho=(1+\delta,1-\delta,1)$, $j=0.6$. (a) $\mathcal R=\eta\varepsilon_\chi/(\omega Z)$ from exact diagonalization for $\delta=0$, $0.4$, and $0.8$ and $25\le\eta\le655360$; stars, $\gamma$ from Eq.~\eqref{eq:gap}. (b) $\sqrt{\det G(s)}/\gamma_0$ from Eq.~\eqref{eq:Gs} for $\delta=0.8$ (light) and $\delta=0$ (dark); symbols, projected Hamiltonian at $\eta=10^8$; gray dashed and dotted lines, the $G_{\rm H}$ and $G$ values for $\delta=0.8$.}
\label{fig:Scov}
\end{figure}

\textit{Convergence to Eq.~\eqref{eq:gap}.}---Figure~\ref{fig:Scov}(a) shows $\mathcal R=\eta\varepsilon_\chi/(\omega Z)$ from exact diagonalization for $\delta=0$, $0.4$, and $0.8$ against $\eta^{-1/3}$. At $\delta=0.8$, where $\Delta_1/\Delta_0=9$, a quadratic fit in $\eta^{-1/3}$ gives an intercept within $0.3\%$ of $\gamma=\gamma_0(1-\delta^2)^{1/2}$, whereas $\sqrt{\det G_{\rm H}}$ exceeds $\gamma$ by $48\%$. The result does not depend on the choice $j=0.6$: at $\eta=40960$, exact diagonalization gives $\mathcal R_{0.8}/\mathcal R_0=0.627$, $0.619$, and $0.616$ for $j=0.3$, $0.6$, and $0.8$, against $0.600$ from $G$ and $0.887$ from $G_{\rm H}$.

\textit{Test of Eq.~\eqref{eq:Gs} in the lowest Landau level.}---We diagonalize the projected Hamiltonian, $(A/2)p_z^2+(r_0/2)z^2+V$, with $V$ of Eq.~\eqref{eq:V} and $Q$, $P$ as operators, for $\rho=(1.8,0.2,1)$ and $(1,1,1)$, $s=0,1,3,10$, and $\eta=10^6$ and $10^8$ [Table~\ref{tab:S3} and Fig.~\ref{fig:Scov}(b)]. Products of $Q$ and $P$ are taken as written; the projection constants of Eq.~\eqref{eq:S4} shift $s$ by $O(\eta^{-1/3})$. The basis is a product of oscillators for $z$ and for the guiding center, the latter squeezed along the principal axes of $G$; enlarging it from $(48,36)$ to $(56,44)$ changes the gaps by less than $2\times10^{-10}$. The residual deviation of $\mathcal R/\gamma(s)$ from unity is the first finite-$\eta$ correction: it falls by a factor of 4.4--4.6 from $\eta=10^6$ to $10^8$, close to $100^{1/3}=4.6$.

\begin{table}[h]
\caption{Projected Hamiltonian away from the triple point, compared with $\gamma(s)=\sqrt{\det G(s)}$ from Eq.~\eqref{eq:Gs}. For \mbox{$\rho=(1.8,0.2,1)$}, $\gamma(0)=0.4593$ is the covariance value and $\gamma_{\rm H}=\sqrt{\det G_{\rm H}}=0.6790$ the second-moment value.}
\label{tab:S3}
\begin{ruledtabular}
\begin{tabular}{cccccc}
$\rho$ & $s$ & $\gamma(s)$ & $\mathcal R$ ($\eta=10^6$) & $\mathcal R$ ($\eta=10^8$) & $\mathcal R/\gamma(s)$ ($\eta=10^8$)\\ \hline
$(1.8,0.2,1)$ & 0 & 0.4593 & 0.4698 & 0.4617 & 1.0052\\
$(1.8,0.2,1)$ & 1 & 0.5441 & 0.5528 & 0.5460 & 1.0035\\
$(1.8,0.2,1)$ & 3 & 0.6113 & 0.6228 & 0.6138 & 1.0041\\
$(1.8,0.2,1)$ & 10 & 0.6608 & 0.6828 & 0.6655 & 1.0072\\
$(1,1,1)$ & 0 & 0.7655 & 0.7724 & 0.7670 & 1.0019\\
$(1,1,1)$ & 1 & 0.7655 & 0.7756 & 0.7677 & 1.0029\\
$(1,1,1)$ & 3 & 0.7655 & 0.7807 & 0.7688 & 1.0043\\
$(1,1,1)$ & 10 & 0.7655 & 0.7926 & 0.7713 & 1.0076
\end{tabular}
\end{ruledtabular}
\end{table}

\textit{Determinant and ground state of the guiding center.}---For positive $\rho_i$, $\bm w^TG\bm w\propto\sum_i\rho_i[(\bm e_i-\ee)\cdot\bm w]^2>0$ for every $\bm w\neq0$, because the three site vectors are not collinear. For three weighted points the scatter matrix obeys
\begin{align}
\det\Big[\sum_i\rho_i(\bm e_i-\ee)(\bm e_i-\ee)^T\Big]=\frac{\rho_0\rho_1\rho_2}{\sum_i\rho_i}\Big(\frac{3\sqrt3}{2}\Big)^2,
\label{eq:S16}
\end{align}
where $3\sqrt3/2$ is twice the area of the triangle with vertices $\bm e_i$; this gives Eq.~\eqref{eq:gap}. The Hamiltonian $(Z/2\eta)\bxi^TG\bxi$ with \mbox{$[Q,P]=\mathrm{i}$} has level spacing $(Z/\eta)\sqrt{\det G}$ and a minimum-uncertainty ground state with covariance $(\gamma/2)G^{-1}$. At $\delta=0.8$ and $\eta=655360$, exact diagonalization of Eq.~\eqref{eq:H} gives $\langle Q^2\rangle=0.409$, $\langle P^2\rangle=0.882$, and $\tfrac12\langle\{Q,P\}\rangle=0.330$, against $0.411$, $0.900$, and $0.346$ from $G^{-1}$. For $\rho=(0.3,1.1,1.6)$, which has no symmetry, exact diagonalization gives $\mathcal R/\gamma=1.038$ at $\eta=655360$, while $G_{\rm H}$ would give $1.27$.

\section*{S4. RESPONSE OF THE GAP TO THE DISTANCE FROM THE TRIPLE POINT}

\textit{Derivation of Eqs.~\eqref{eq:ratio} and \eqref{eq:ac}.}---To leading order $\varepsilon_\chi/\omega=(Z/\eta)\sqrt{\det G(s)}$ with $Z=\ell_z^2\eta^{1/3}f(s)$. With Eq.~\eqref{eq:Gs} and the rank-one identity of Sec.~S3, $\det G(s)=\det G\,\{1+a[1-\varphi(s)]\}$, which gives Eq.~\eqref{eq:ratio}. Its logarithmic derivative at $s=0$ is $f'(0)/f(0)-\tfrac a2\varphi'(0)$; with $\varphi'(0)=-1/[6f(0)]$ from Eq.~\eqref{eq:S14b} and $f'(0)=-0.0690205$ this is Eq.~\eqref{eq:ac}, $a_c=12|f'(0)|=0.828246$. The threshold involves only the quartic oscillator; the system enters only through $a$. For $s\gg1$, $\varphi\simeq(11/3)s^{-3/2}$ and $f\simeq1/(2\sqrt s)$, so the ratio decreases as $s^{-1/2}$ and has a maximum for every $a>a_c$. Table~\ref{tab:S4} lists the maximum for the values of $\rho$ used in the main text, and Fig.~\ref{fig:3}(a) shows its position $s_{\max}(a)$.

\begin{table}[h]
\caption{Leading-order response, Eq.~\eqref{eq:ratio}, for $\rho=(1+\delta,1-\delta,1)$: parameter $a$, maximum gain of the gap and its position, the critical variance there, and the point $s_c$ of Eq.~\eqref{eq:sc}. The last row is $\delta=(147/173)^{1/2}$, for which $a=49/13$ exactly.}
\label{tab:S4}
\begin{ruledtabular}
\begin{tabular}{cccccc}
$\delta$ & $a$ & $\max_s\varepsilon_\chi(s)/\varepsilon_\chi(0)$ & $s_{\max}$ & $Z(s_{\max})/Z(0)$ & $s_c$\\ \hline
0 & 0 & 1 (at $s=0$) & 0 & 1 & ---\\
0.8 & 1.185 & 1.0155 & 0.434 & 0.925 & $-1.163$\\
$\sqrt3/2$ & 2 & 1.0953 & 0.982 & 0.847 & $-0.805$\\
0.9218 & 49/13 & 1.2915 & 1.481 & 0.789 & $-0.4846$
\end{tabular}
\end{ruledtabular}
\end{table}

\textit{Path at fixed local quartic coefficients.}---Equation~\eqref{eq:ratio} assumes that $u$ and the couplings to $\bxi$ stay fixed while $r_0$ changes. Fixing $g$ fixes the local quartic coefficients $g^4\rho_i/(4\eta)$ of the $x_i$. We therefore keep $g^2=A_*(j{=}0.6)$ and set $r_0$ through $(j,\theta)$: $A=1-j\cos\theta=g^2+r_0$ and $r_\chi=0$ give $j\cos\theta=1-g^2-r_0$ and $(j\sin\theta)^2=\tfrac43C(C-g^2)$ with $C=1+\tfrac12j\cos\theta$. Along this path $A$ and $\ell^2$ still change, $d\ln A/dr_0=1.18$ and $d\ln\ell^2/dr_0=0.84$, so the prefactor $g^{8/3}\ell^2A^{1/3}$ of the leading gap rises by $2\%$ at $r_0=0.016$, which is $s=1.5$ at $\eta=100$. Changing $g$ instead, at fixed $\theta$ with $j$ adjusted to keep $r_\chi=0$, gives $d\ln g^4/dr_0=-4.1$ and $d\ln\ell^2/dr_0=-4.3$, which lower this prefactor by $10\%$ at the same $r_0$. At $\eta=100$ and $\delta\simeq0.922$ exact diagonalization of Eq.~\eqref{eq:H} on this second path gives $\varepsilon_\chi(1.5)/\varepsilon_\chi(0)=0.994$ instead of $1.147$, with nearly the same $Z(1.5)/Z(0)=0.756$ instead of $0.755$: at small $\eta$ the choice of the tuning parameter decides whether the effect is seen. Both changes of the prefactor vanish as $\eta^{-2/3}$ at fixed $s$.

\textit{Numerical results.}---Along this path we diagonalize Eq.~\eqref{eq:H}, with all eight atomic states, at $\eta=25$ and $100$, and $H_{\rm eff}$ at $\eta=10^4$, $10^6$, and $10^8$, for $\delta=0$, $0.8$, $\sqrt3/2$ (only $H_{\rm eff}$), and $0.922$ and for $s=0.5$, $1$, $1.5$, $2$, and $3$; the full table of $\varepsilon_\chi(s)/\varepsilon_\chi(0)$ and $Z(s)/Z(0)$ is in the Zenodo archive. At $\eta=100$ and $s=1.5$, Eq.~\eqref{eq:H} gives $\varepsilon_\chi(s)/\varepsilon_\chi(0)=1.1475$, $1.0162$, and $0.9425$ for $\delta\simeq0.922$, $0.8$, and $0$, with $Z(s)/Z(0)=0.7550$, $0.7932$, and $0.8293$. At $\eta=10^8$, $H_{\rm eff}$ deviates from Eq.~\eqref{eq:ratio} by at most $0.73\%$ in the gap ratio and from $f(s)/f(0)$ by at most $0.15\%$ in the variance ratio, for all four values of $\delta$ and all five values of $s$. The effective Hamiltonian $H_{\rm eff}$ is diagonalized in a product basis of a squeezed oscillator for $z$, the guiding-center oscillator $c_+$, and the cyclotron oscillator $c_-$ (cutoffs 36, up to 260, and 8); at the triple point it reproduces the gaps and variances obtained from Eq.~\eqref{eq:H} to $0.01\%$ at $\eta=655360$. For this path, Eq.~\eqref{eq:H} is diagonalized in the Fock basis with total photon number up to $N_{\rm ph}=30$ ($\eta=25$) and $40$ ($\eta=100$); at the triple point and $\eta=25$ it reproduces the product-basis results of Sec.~S8 to five significant digits, and reducing $N_{\rm ph}$ to $34$ at $\eta=100$, $\delta\simeq0.922$, $s=1.5$ changes the gap by $5\times10^{-5}$ in relative terms.

\section*{S5. OVERCOMPENSATION AND THE SECONDARY PHASE TRANSITION}

\textit{Instability.}---For $s<0$ the potential $y^4+sy^2$ has two wells. As $t=-s\to\infty$, the ground state and the first odd state become nearly degenerate, with splitting $\mathcal E_1-\mathcal E_0$, and the matrix element of $y^3$ between them approaches $(t/2)^{3/2}$; hence $\mathcal X_3\simeq t^3/[4(\mathcal E_1-\mathcal E_0)]$, $f\simeq t/2$, and $\varphi\simeq2t^2/[3(\mathcal E_1-\mathcal E_0)]\to\infty$. Since $\varphi(0)=1$ and $\varphi$ is continuous, Eq.~\eqref{eq:sc} has a solution $s_c<0$ for every $a>0$; Table~\ref{tab:S4} lists examples, and $\varphi(-1)=1.675$, $\varphi(-2)=3.228$, $\varphi(-3)=7.608$.

\textit{Classical limit of the guiding center.}---With $z=\ell_z\eta^{1/6}y$ and $\bm q=\ell\bxi/(\ell_z\eta^{1/6})$, the projected Hamiltonian in units of $A/(2\ell_z^2\eta^{1/3})$ reads $p_y^2+sy^2+\bar\rho^{-1}\sum_i\tfrac{\rho_i}{3}(y+\bm e_i\cdot\bm q)^4$, with $[q_1,q_2]=\mathrm{i}\hbar_{\rm eff}$ and $\hbar_{\rm eff}=\ell^2/(\ell_z^2\eta^{1/3})$, which is $1.1172\,\eta^{-1/3}$ at $j=0.6$. The rescaling is exact; $\eta$ enters only through $\hbar_{\rm eff}$. As $\hbar_{\rm eff}\to0$ the ground-state energy is the minimum over $\bm q$ of the adiabatic potential $V_{\rm ad}(\bm q)$, the ground-state energy of the $y$ problem at fixed $\bm q$, while the $y$ problem stays fully quantum. Expanding $V_{\rm ad}$ about the origin gives $V_{\rm ad}=\tfrac12\bm q^T\mathcal K\bm q+\dots$ with $\mathcal K\propto G(s)$; at $s_c$ the unstable direction is $\hat{\bm n}\propto G^{-1}\ee$.

\textit{Order.}---The quartic form is even under $(z,\bxi)\to(-z,-\bxi)$, so $V_{\rm ad}(\bm q)=V_{\rm ad}(-\bm q)$, and for $s<s_c$ the minimum of $V_{\rm ad}$ lies at $\bm q\neq0$: a phase transition that breaks the $\mathbb{Z}_2$ parity must occur at some $s\ge s_c$. Along $\hat{\bm n}$,
\begin{align}
V_{\rm ad}=u_2(s)\,q_{\rm ord}^2+u_4\,q_{\rm ord}^4+\cdots,\qquad u_2(s_c)=0.
\label{eq:Vad}
\end{align}
If $u_4(s_c)>0$ and $V_{\rm ad}$ has no other minimum below $V_{\rm ad}(0)$ at $s\ge s_c$, the minimum leaves the origin continuously at $s_c$, a second-order transition. If $u_4<0$, or a distant minimum crosses $V_{\rm ad}(0)$ first, the transition is of first order and occurs at some $s>s_c$; then Eq.~\eqref{eq:sc} gives the spinodal. This is the standard Landau argument; the order of the transition is not fixed by $a$.

For the Rabi ring we compute $u_4(s_c)$ for $80$ choices of $\rho=(1,\rho_1,\rho_2)$ with $0.05\le\rho_{1,2}\le2$ ($0.011\le\beta\le0.863$, $\beta=a/(1+a)$), and search $V_{\rm ad}$ for distant minima at $s=s_c+0.03$ from 36 starting points. In all cases $u_4\ge0.097$, and no competing minimum is found within the values of $\rho$ and the search range examined, so the transition is of second order; for $\delta\simeq0.922$, $u_4=0.170$.

For sums of local quartic terms with generic site vectors, $u_4$ is often small and can be marginally negative, and nonlocal quartic couplings can make the transition clearly first order. An example, whose quartic form is not a sum of fourth powers of linear forms, is the general effective Hamiltonian of Sec.~S6 with quartic part $z^4+z^3Q+z^2Q^2+\tfrac32zQ^3+\tfrac34Q^4+z^2P^2+P^4$, which has $a=3/5$ and $s_c=-1.7343$ but $u_4(s_c)=-0.109$; its global minimum jumps from $\bm q=0$ to $|\bm q|=1.519$ at $s=-1.4658$, and exact diagonalization at $\hbar_{\rm gen}=10^{-2}$ and $3\times10^{-3}$ (normalization of Sec.~S6) shows the Binder cumulant jumping from $0.66$ to $0$ between $s=-1.475$ and $-1.450$ without softening at $s_c$.

\textit{Scaling at the second-order transition.}---Near $s_c$ the guiding center is described by
\begin{align}
h_{\rm ord}=\alpha_{\rm kin}\,p_{\rm ord}^2+u_2'(s_c)\,(s-s_c)\,q_{\rm ord}^2+u_4\,q_{\rm ord}^4,\qquad[q_{\rm ord},p_{\rm ord}]=\mathrm{i}\hbar_{\rm eff},
\label{eq:hord}
\end{align}
in units of $A/(2\ell_z^2\eta^{1/3})$, with $\alpha_{\rm kin}>0$ the curvature of $V_{\rm ad}$ along the direction conjugate to $q_{\rm ord}$. Equation~\eqref{eq:hord} is the low-energy effective Hamiltonian of the Rabi model at its superradiant transition~\cite{Hwang2015}, with the uniform mode in the role of the atom and the guiding center in the role of the cavity, and with the ratio of their bare frequency scales $\propto\hbar_{\rm eff}^{-1}\propto\eta^{1/3}$; at $s_c$ itself the ratio of excitation energies is $\varepsilon_z/\varepsilon_\chi\propto\eta^{4/9}$. The transition is defined in the joint limit $\eta\to\infty$ at fixed $s$, and the finite-$\eta$ Binder curves below show its finite-size rounding.

In the Rabi model the gap, the amplitude, and the critical region scale as $\eta_R^{-1/3}$, $\eta_R^{1/6}$, and $\eta_R^{-2/3}$ in the frequency ratio $\eta_R$, here $\eta_R\propto\hbar_{\rm eff}^{-1}\propto\eta^{1/3}$. This gives, at $s_c$, $\varepsilon_\chi\propto\eta^{-2/3}\eta^{-1/9}=\eta^{-7/9}$, $\langle q_{\rm ord}^2\rangle^{1/2}\propto\hbar_{\rm eff}^{1/3}$, and a critical region $|s-s_c|\propto\eta^{-2/9}$, that is, $\propto\eta^{-8/9}$ in $r_0$.

For $s<s_c$ the guiding center sits in one of two minima at $|\bm q|=O(1)$, and the splitting between the two lowest states of opposite parity decays as $\exp(-\mathcal S_{\rm tun}/\hbar_{\rm eff})=\exp(-\mathcal S_{\rm tun}'\eta^{1/3})$. The same exponent $7/9$ appears in the quartic crossover near the endpoint $a\to\infty$ of the line $s_c(a)$ (Sec.~S7, $\kappa=\rho_2\eta^{2/3}\ll1$); at the endpoint itself, $\rho_2=0$, the gap vanishes identically.

\textit{Numerical test.}---Figure~\ref{fig:3}(b) shows the Binder cumulant $U_4=1-\langle q_{\rm ord}^4\rangle/(3\langle q_{\rm ord}^2\rangle^2)$ of the order parameter $q_{\rm ord}=\hat{\bm n}\cdot\bm q$ in the ground state of $H_{\rm eff}$ for $\delta\simeq0.922$. It tends to $0$ for $s>s_c$ and to $2/3$ for $s<s_c$ as $\eta$ grows. The crossings of $U_4$ for successive pairs of $\eta$, from $(10^3,10^4)$ to $(10^7,10^8)$, lie at $s=-0.456$, $-0.475$, $-0.480$, $-0.482$, and $-0.483$, approaching $s_c=-0.4846$. For the general effective Hamiltonian of Sec.~S6, computed down to $\hbar_{\rm gen}=3\times10^{-4}$, the crossings of $\varepsilon_\chi/\hbar_{\rm gen}^{4/3}$ lie within $1.3\times10^{-3}$ of $s_c$ for every coupling set tested, and data at different $\hbar_{\rm gen}$ collapse when plotted as $\varepsilon_\chi/\hbar_{\rm gen}^{4/3}$ against $(s-s_c)\hbar_{\rm gen}^{-2/3}$. For the Kerr set of Sec.~S6 ($a=49/13$), the splitting of the two lowest states at $s=-0.7$ decreases as $\exp(-\mathcal S_{\rm tun}/\hbar_{\rm gen})$ with a single $\mathcal S_{\rm tun}\simeq0.038$ for $\hbar_{\rm gen}=0.02$, $0.006$, and $0.002$.

\section*{S6. GENERAL EFFECTIVE HAMILTONIAN AND OTHER REALIZATIONS}

\textit{Definition.}---Consider a critical coordinate $z$ with kinetic energy $(A/2)p_z^2$ and a zero-frequency mode with quadratures $\bxi=(Q,P)^T$, $[Q,P]=\mathrm{i}$, coupled through a general quartic form,
\begin{align}
H_{\rm gen}=\frac A2p_z^2+\frac{r_0}{2}z^2+\frac1\Lambda\Big[uz^4+z^3\,\bm c\cdot\bxi+\frac12z^2\,\bxi^TB\bxi+zT_3(\bxi)+T_4(\bxi)\Big],
\label{eq:S20}
\end{align}
with $T_{3,4}$ homogeneous of degree three and four and the whole form bounded below; $s=Mr_0\Lambda^{2/3}$ is defined as in Eq.~\eqref{eq:S6} with $\Lambda$ in place of $\eta$. We assume that the state $\langle\bxi\rangle=0$ is stable at $s=0$, that is, $B\succ0$ and $B-\tfrac{3}{4u}\bm c\bm c^T\succ0$, equivalently $0\le\beta<1$ with $\beta$ defined below; boundedness of the quartic form alone does not ensure this. The rescaling $z=(A\Lambda/2u)^{1/6}y$, $\bxi=(A\Lambda/2u)^{1/6}\bm q_{\rm gen}$ is exact and leaves a single small parameter $\hbar_{\rm gen}=(2u/A\Lambda)^{1/3}$ in $[q_{{\rm gen},1},q_{{\rm gen},2}]=\mathrm{i}\hbar_{\rm gen}$; we denote this normalization by $\hbar_{\rm gen}$ to distinguish it from $\hbar_{\rm eff}$ of Sec.~S5. For the ring, $\hbar_{\rm gen}=\hbar_{\rm eff}/\ell^2$, because the coordinate of Sec.~S5 contains an extra factor of $\ell$.

\textit{Leading order.}---To leading order the curvature of the adiabatic potential is $\mathcal K(s)\propto f(s)\{B-[3/(4u)]\varphi(s)\bm c\bm c^T\}$, the compensation again removing a rank-one part. With the bounded plotting coordinate $\beta=a/(1+a)=[3/(4u)]\,\bm c^TB^{-1}\bm c$, Eqs.~\eqref{eq:ratio} and \eqref{eq:ac} and the instability condition, Eq.~\eqref{eq:sc}, follow unchanged; whether the instability is a second-order transition is decided as in Sec.~S5. The terms $T_3$ and $T_4$ and the direction of $\bm c$ relative to $B$ enter only the finite-$\hbar_{\rm gen}$ corrections, the squeezing axes, and the order of the secondary transition. For the ring, $\bm c\propto\ee$ and $B\propto G_{\rm H}$. The exactly flat direction of Sec.~S7 at $\beta=1$ needs more than the general effective Hamiltonian: there a magnetic translation leaves every nonlinear coordinate unchanged, whereas in general $T_4$ can still confine $\bxi$ at $\beta=1$.

\begin{figure}[h]
\includegraphics[width=0.92\textwidth]{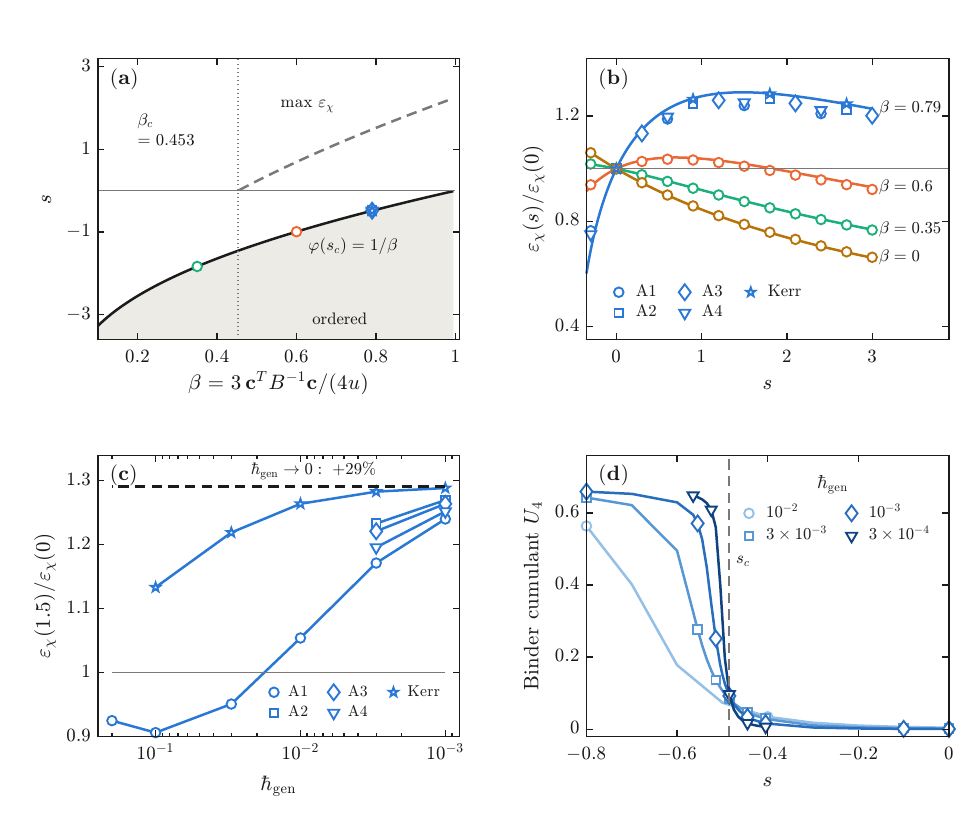}
\caption{General effective Hamiltonian, Eq.~\eqref{eq:S20}, for eight coupling sets. (a) Phase diagram in $\beta$ and $s$: solid, Eq.~\eqref{eq:sc}, written as $\varphi(s_c)=1/\beta$; dashed, maximum of the gap; dotted, $\beta_c$; shaded, ordered phase; symbols, crossings of $\varepsilon_\chi/\hbar_{\rm gen}^{4/3}$ between $\hbar_{\rm gen}=10^{-3}$ and $3\times10^{-4}$. (b) $\varepsilon_\chi(s)/\varepsilon_\chi(0)$ at $\hbar_{\rm gen}=10^{-3}$; lines, Eq.~\eqref{eq:ratio}. (c) Approach of $\varepsilon_\chi(1.5)/\varepsilon_\chi(0)$ to the leading value (dashed) for the four random $\beta=0.790$ sets (A1--A4) and the Kerr set. (d) Binder cumulant $U_4$ of the order parameter for the Kerr set.}
\label{fig:Snormal}
\end{figure}

\textit{Test with random coupling sets.}---We diagonalize Eq.~\eqref{eq:S20} exactly in the scaled form for $3\times10^{-4}\le\hbar_{\rm gen}\le0.2$ for eight coupling sets: four with $\beta=0.790$ ($a=49/13$) but independently drawn $B$ (eigenvalues between 3 and 15, random axes), directions of $\bm c$, and $T_{3,4}$ (sums of powers of random linear forms); one each with $\beta=0.60$ ($a=1.5$) and $0.35$ ($a=0.54$); a control with $\bm c=0$; and a set from a parametrically driven Kerr oscillator coupled to a resonant mode (below). The basis and convergence are as in Sec.~S8. Figure~\ref{fig:Snormal} shows the results. The crossings of $\varepsilon_\chi/\hbar_{\rm gen}^{4/3}$ between $\hbar_{\rm gen}=10^{-3}$ and $3\times10^{-4}$ are $-0.4834$ to $-0.4839$ for the four $\beta=0.790$ sets and $-0.4841$ for the Kerr set, against $s_c=-0.4846$; $-0.9912$ for $\beta=0.60$ ($s_c=-0.9917$); and $-1.8323$ for $\beta=0.35$ ($s_c=-1.8324$). The control shows no transition.

At $\hbar_{\rm gen}=10^{-3}$, $\varepsilon_\chi(1.5)/\varepsilon_\chi(0)$ is $1.240$--$1.270$ for the random $\beta=0.790$ sets and $1.289$ for the Kerr set, against $1.292$ from Eq.~\eqref{eq:ratio}, and $1.010$, $0.875$, and $0.788$ for $\beta=0.60$, $0.35$, and $0$, against $1.018$, $0.877$, and $0.787$. The approach to the leading curve depends on the coupling set [Fig.~\ref{fig:Snormal}(c)]: the residual is set by the zero-point motion of $\bxi$ acting through $T_4$ and $T_3$ and is enhanced by $1/(1-\beta)$.

\textit{Zero-frequency mode from resonance.}---The flux is one way to obtain a zero-frequency mode. Another is a cavity mode (mode 2) held at zero detuning in the frame of a two-photon drive that brings a Kerr parametric oscillator (mode 1) to its threshold, the critical mode~\cite{PradoReynoso2025}. If the two modes share a Kerr element of strength $K_{12}$ with participations $\tau_1$ and $\tau_2$, while mode 1 has its own Kerr coefficient $K_1$ and mode 2 a self-Kerr coefficient $K_2$, the quartic form after the rotating-wave approximation, with the gapped quadrature of mode 1 dropped, is
\begin{align}
V_4=K_1z^4+K_{12}\big[(\tau_1z+\tau_2Q)^2+\tau_2^2P^2\big]^2+K_2\big(Q^2+P^2\big)^2,
\end{align}
so that $u=K_1+K_{12}\tau_1^4$, $\bm c=(4K_{12}\tau_1^3\tau_2,0)$, and $B=K_{12}\tau_1^2\tau_2^2\,\mathrm{diag}(12,4)$, which gives $\beta=K_{12}\tau_1^4/(K_1+K_{12}\tau_1^4)$, the share of the critical Kerr nonlinearity carried by the shared element, independent of $\tau_2$ and of $K_2$. In the rescaled quadratures $(\tau_2/\tau_1)(Q,P)$ the shared term depends only on $\tau_1$, so the ratio $\tau_2/\tau_1$ enters only through $\hbar_{\rm gen}\propto(\tau_2/\tau_1)^2\Lambda^{-1/3}$ and through the rescaled self-Kerr coefficient $K_2(\tau_1/\tau_2)^4$ in $T_4$. The Kerr set of Fig.~\ref{fig:Snormal} has $K_1/K_{12}=13/49$, $K_2=K_{12}$, $\tau_1=1$, and $\tau_2=0.7$, so that $\beta=49/62$.

\textit{Collective-spin triangle.}---The triangular coupled-top model of Ref.~\cite{Duan2023} with an added Dzyaloshinskii--Moriya interaction, which plays the role of the hopping phase~\cite{Padilla2022}, is
\begin{align}
H_{\rm sp}=\sum_{i=0}^{2}\Big[\Omega S_i^z-\frac{J_{\rm sp}}{S}S_i^xS_{i+1}^x+\frac{D_{\rm sp}}{S}\big(S_i^xS_{i+1}^y-S_i^yS_{i+1}^x\big)\Big].
\end{align}
In the Holstein--Primakoff representation about the downward-polarized state, with $S_i^x\simeq\sqrt S x_i$ and $S_i^y\simeq-\sqrt Sp_i$ to leading order, its quadratic part is
\begin{align}
h_{{\rm sp},2}=\frac\Omega2p_z^2+\frac{\Omega-2J_{\rm sp}}{2}z^2+\frac\Omega2(p_X^2+p_Y^2)+\frac{\Omega+J_{\rm sp}}{2}R^2-\sqrt3D_{\rm sp}L_z,
\end{align}
and the uniform and chiral boundaries meet at $J_{\rm sp}=\Omega/2$, $D_{\rm sp}=\Omega/\sqrt2$. Projection onto the lowest Landau level, with Weyl ordering of the $1/S$ terms, gives Eq.~\eqref{eq:S20} at leading order with $\Lambda=S$ ($N=2S$ atoms per site). For identical sites the $C_3$ symmetry forces $\bm c=0$, so $a=0$; a nonzero $a$ requires inequivalent sites. The diagonal part has the form of Eq.~\eqref{eq:S5} with $A=1$, $u=1/24$, $\gamma_{\rm sp}=\sqrt6/8$ in place of $\gamma_0$, and $v=17/96$. Two-thirds of $\gamma_{\rm sp}$ come from a $z^2L_z$ coupling, absent in the ring, which gives $z^2n$ rather than $z^2(n+1)$ in the lowest Landau level. At the spin triple point, in units of $\Omega$, $\varepsilon_\chi=0.2538\,S^{-2/3}-0.2213\,S^{-1}$, $Z=0.8288\,S^{1/3}+0.2451$, and $\mathcal R=\sqrt6/8-0.3575\,S^{-1/3}$, so the first correction has the opposite sign from that in the ring. Exact diagonalization with a total-excitation cutoff gives $\mathcal R/\gamma_{\rm sp}=0.661$, $0.851$, and $0.939$ at $S=25$, $400$, and $6400$, and the two-term gap agrees to $0.07\%$ at $S=6400$. Here $Z$ is the variance of the Holstein--Primakoff coordinate $z=\sum_ix_i/\sqrt3$; that of $\sum_iS_i^x/\sqrt{3S}$ differs at $O(S^{-1/3})$ and changes $\mathcal R$ only at $O(S^{-2/3})$.

\textit{Rotating trap.}---For a particle in a trap rotating at its radial frequency, with no axial confinement, the guiding center of the radial motion has zero frequency in the rotating frame, and the axial coordinate $z$, confined only by the quartic anharmonicity, plays the role of the critical mode~\cite{Fetter2009}. An anharmonicity symmetric about the rotation axis, such as one $\propto(R^2+z^2)^2$ with $R$ the radial coordinate, contains only $z^4$, $z^2R^2$, and $R^4$. It has no term odd in the radial coordinates and therefore realizes Eq.~\eqref{eq:S20} with $\bm c=0$ and $a=0$; a nonzero $a$ requires a co-rotating anharmonicity that breaks the axial symmetry, for example one containing $z^3x$.

\section*{S7. ONE ATOM WITHOUT NONLINEARITY}

\textit{Exact magnetic-translation symmetry.}---The limit $\rho_2\to0$ at fixed $g$ means $\Delta_2\to\infty$ with $\lambda_2\propto\sqrt{\Delta_2}$; atom 2 then adjusts to $x_2$ exactly and contributes only $-g^2x_2^2/2$. Let $\bm\nu=(-1/2,\sqrt3/2)^T$, so that $\bm e_0\cdot\bm\nu=\bm e_1\cdot\bm\nu=-1/2$ and $\bm e_2\cdot\bm\nu=1$. Rotate $(Q,P)$ canonically so that $Q_\parallel=\bm\nu\cdot\bxi$. Shifting $Q_\parallel\to Q_\parallel+t$ together with $z\to z+\ell t/2$ leaves $z+\ell\,\bm e_i\cdot\bxi$ unchanged for $i=0,1$. In the photon coordinates of Eq.~\eqref{eq:h2} the same operation is a magnetic translation of $(X,Y)$ by $(\delta_X,\delta_Y)$, with $p_X\to p_X-(b/C)\delta_Y$ and $p_Y\to p_Y+(b/C)\delta_X$, combined with a shift of $z$. It leaves $p_z$ and the covariant momenta in Eq.~\eqref{eq:h2} unchanged, so it commutes with $h_2$ at $r_0=r_\chi=0$, and it leaves $x_0$ and $x_1$ unchanged. The nonlinear terms of atoms 0 and 1 depend only on $x_0$ and $x_1$. The magnetic translation is therefore a symmetry of the Hamiltonian projected onto the lower atomic levels at every order in $1/\eta$, and the energy of the guiding center is exactly flat along $\bm\nu$, with no normalizable ground state. Because the atoms couple to the photons only through $x_i$, the argument does not use the projection of atoms 0 and 1 onto their lower level: the magnetic translation commutes with Eq.~\eqref{eq:H} once atom 2 is replaced by its limit $-g^2x_2^2/2$. Numerically, the ground-state energy of $z$ is independent of $t$ to machine precision for $\rho=(1.5,1.5,0)$ and $(1,2,0)$. The same construction applies whenever all sites with $\rho_i>0$ satisfy $\bm e_i\cdot\bm\nu=\mathrm{const}$. In the language of Sec.~S3, $\det G=0$ and $a=\infty$: the compensation removes the whole confinement along $\bm\nu$.

\textit{Crossover.}---For small $\rho_2>0$ and $\rho_0=\rho_1=1$, the confinement along $\bm\nu$ comes from atom 2 alone. After the shift of $z$ of Sec.~S3, the coordinate $x_2$ of atom 2 contains $\ell(\bm e_2-\ee)\cdot\bm\nu\,Q_\parallel=[3\ell/(2+\rho_2)]Q_\parallel$, so the guiding-center Hamiltonian is
\begin{align}
h_{\rm gc}=\frac{Z_{\rm f}}{2\eta}\bxi^TG\bxi+wQ_\parallel^4,\qquad w=\frac{g^4\rho_2\ell^4}{36\eta}\Big(\frac{3}{2+\rho_2}\Big)^4,\qquad Z_{\rm f}=f(0)\Big(\frac{A\eta}{2u}\Big)^{1/3},\qquad u=\frac{g^4(2+\rho_2)}{36}.
\end{align}
Here $Z_{\rm f}$ is the variance of $z$ at fixed guiding center, that is, of $z_{\rm f}=z+\ell\,\ee\cdot\bxi\simeq z-(\ell/2)Q_\parallel$; the total variance $\langle z^2\rangle=Z_{\rm f}+(\ell^2/4)\langle Q_\parallel^2\rangle$ differs from it at leading order when $\kappa=\rho_2\eta^{2/3}\lesssim1$ (by $42\%$ at $\eta=10^6$, $\kappa=0.1$). Equation~\eqref{eq:gap}, with $Z$ the total variance, therefore applies as $\eta\to\infty$ at fixed $\rho_2>0$, whereas at fixed $\kappa$ the gap follows from $h_{\rm gc}$. At leading order $\eta\varepsilon_\chi/\omega=\Phi(\kappa)$, with corrections of relative order $\rho_2$. For $\kappa\gg1$, $\Phi\to0.747\,\kappa^{1/2}$, the small-$\rho_2$ limit of Eq.~\eqref{eq:gap}; for $\kappa\ll1$ the quartic term dominates and, with the gap 2.7393 of $p^2+y^4$,
\begin{align}
\frac{\varepsilon_\chi}{\omega}\simeq2.7393\Big(\frac{\gamma_0Z_{\rm f}}{2\eta}\Big)^{2/3}w^{1/3}=0.960\,\rho_2^{1/3}\eta^{-7/9}\qquad(j=0.6).
\end{align}
The projected Hamiltonian, Eq.~\eqref{eq:V} with $Q$, $P$ as operators, agrees with $\Phi(\kappa)$ to within $0.6\%$ for $0.1\le\kappa\le10$ at $\eta=10^5$ and $10^6$. At $\kappa=100$ it lies below $\Phi$ by $11.8\%$, $3.0\%$, and $0.6\%$ at $\eta=10^4$, $10^5$, and $10^6$; most of this is the finite-$\rho_2$ correction already contained in $h_{\rm gc}$, and the remainder, from treating $\bxi$ as an operator, is at most $0.2\%$ at $\eta=10^6$. A small distance $\delta r$ from either phase boundary adds a quadratic confinement $\propto\delta r$ along $\bm\nu$, so the quartic regime requires $|\delta r|\ll\rho_2^{2/3}\eta^{-8/9}$.

\section*{S8. NUMERICAL METHODS AND CONVERGENCE}

\textit{Exact diagonalization of Eq.~\eqref{eq:H}.}---Equation~\eqref{eq:H} is diagonalized with all eight atomic states in a product basis of one squeezed oscillator for $z$ and two circular oscillators. The uniform basis length follows the quartic scale $(6A\eta/g^4)^{1/6}$, and the length of the two circular oscillators is $\ell$. Polynomial operators are formed in a space padded by four states per oscillator and then truncated. Inhomogeneous atomic frequencies break the $C_3$ symmetry of the ring, so only total parity is used. The lowest states of each parity are obtained by shift-invert Lanczos iteration; the residual norms of the ground and chiral states are below $2\times10^{-7}$ times the gap, and those of all requested states below $2\times10^{-6}$ times the gap. The chiral excitation is the pole with the largest weight in $W|0\rangle$; at every point it is also the lowest excitation overall and carries at least $97\%$ of the retained low-energy weight of $W|0\rangle$. The uniform excitation $\varepsilon_z$ is the pole with the largest weight in $z|0\rangle$. For identical atoms at $\eta\ge10^7$ (Table~\ref{tab:S1}), the Hamiltonian is diagonalized in $C_3$ (momentum) blocks, and the final energies are evaluated as Rayleigh quotients with extended-precision accumulation, because the atomic energies exceed the gap by a factor of up to $10^{17}$.

Photon cutoffs $(N_z,N_+,N_-)=(22,12,6)$ are compared with $(26,16,8)$ for $\eta\ge160$ and with $(28,16,8)$ for $\eta\le100$. For $\eta=655360$ at $\delta=0.8$ the comparison is instead between $(26,16,8)$ and $(30,20,10)$, which changes the gap by $5.8\times10^{-6}$. Over all points at the triple point, the largest relative change of a gap is $2.5\times10^{-4}$. The covariances $\langle Q^2\rangle$, $\langle P^2\rangle$, and $\langle\{Q,P\}\rangle/2$ are evaluated with the $c_+$ operators of the circular basis. The points of Sec.~S4 away from the triple point at $\eta=25$ and $100$ use an independent Fock-basis implementation with Lanczos iteration in each parity sector.

\textit{Low-energy effective Hamiltonian $H_{\rm eff}$ and the control without compensation.}---The effective Hamiltonian $H_{\rm eff}=h_2+g^4\sum_i\rho_ix_i^4/(4\eta)$ is diagonalized at the same points with cutoffs up to $(30,22,10)$; it agrees with the gaps of Eq.~\eqref{eq:H} to $0.02\%$ at $\eta=655360$, and an independent implementation reproduces it. The implementation of Sec.~S4, with a larger guiding-center basis, agrees to $0.01\%$. For the control in Fig.~\ref{fig:2}(a), we remove from $\sum_i\rho_ix_i^4$ the terms cubic in $z$ and linear in $(X,Y)$, $4\sum_i\rho_i(z/\sqrt3)^3\sqrt{2/3}\,(X\cos\alpha_i+Y\sin\alpha_i)$, which produce the compensation; for identical atoms these terms cancel. At $\eta=25$, $100$, and $2560$, $\mathcal R_{0.8}/\mathcal R_0$ is $0.645$, $0.635$, and $0.631$ for Eq.~\eqref{eq:H}, $0.620$, $0.626$, and $0.630$ for $H_{\rm eff}$, and $0.807$, $0.830$, and $0.875$ for $H_{\rm eff}$ without these terms, against the leading-order values $0.600$ for $G$ and $0.887$ for $G_{\rm H}$; cutoffs $(30,20,8)\to(36,26,10)$ change the gaps by less than $10^{-7}$. Away from the triple point at $\eta=25$, $H_{\rm eff}$ gives $\mathcal R_{0.8}/\mathcal R_0=0.620$, $0.677$, and $0.741$ for $s=0$, $1$, and $3$, against $0.600$, $0.711$, and $0.799$ from Eq.~\eqref{eq:Gs}; at $\eta=2560$ it gives $0.630$, $0.710$, and $0.783$.

\textit{Projected Hamiltonian for one weakly nonlinear atom.}---For the crossover of Sec.~S7, the guiding-center oscillator is squeezed to the quartic scale along $\bm\nu$ and the $z$ oscillator is scaled to the quartic length of $z$. Cutoffs $(60,40)$ and $(80,54)$ are compared; at $\eta=10^6$ and $\kappa\le1$ we also use $(80,90)$, which changes the gaps by less than $10^{-6}$, and the remaining basis uncertainty, from the $z$ cutoff, is below $0.5\%$.

\textit{General effective Hamiltonian.}---For Secs.~S5 and S6 the scaled Hamiltonian is diagonalized in a product basis of an oscillator for $y$ (36 states) and a guiding-center oscillator of length $\sqrt{\hbar_{\rm eff}}$ in Sec.~S5 or $\sqrt{\hbar_{\rm gen}}$ in Sec.~S6 squeezed to the curvature at $s=0$ (up to 1400 states); the lowest states are found by shift-invert Lanczos iteration in each parity sector. Enlarging the $y$ basis to 48 or the guiding-center basis by a factor of 1.5--2 changes the gaps by less than $10^{-8}$.

\end{document}